\documentclass[%
reprint,amsmath,amssymb,aps,prl,superscriptaddress,
]{revtex4-1} 

\usepackage{graphicx}
\usepackage{dcolumn}
\usepackage{bm}
\usepackage{xfrac} 

\usepackage{mathtools}
\usepackage{hyperref}
\usepackage{fixltx2e}
\usepackage{amsmath,amssymb}
\usepackage{graphicx}
\usepackage[justification=RaggedRight, textfont={sf, small}]{caption}
\usepackage[export]{adjustbox}

\begin{document}

\title{Low-frequency charge noise in Si/SiGe quantum dots}

\author{Elliot J. Connors}
\affiliation{%
Department of Physics and Astronomy, University of Rochester, Rochester, NY 14627
}
\author{JJ Nelson}
\affiliation{%
Department of Physics and Astronomy, University of Rochester, Rochester, NY 14627
}
\author{Haifeng Qiao}
\affiliation{%
Department of Physics and Astronomy, University of Rochester, Rochester, NY 14627
}
\author{Lisa F. Edge}
\affiliation{%
HRL Laboratories LLC, 3011 Malibu Canyon Road, Malibu, California 90265, USA
}
\author{John M. Nichol}
\email{jnich10@ur.rochester.edu}
\affiliation{%
Department of Physics and Astronomy, University of Rochester, Rochester, NY 14627
}


\begin{abstract}
Electron spins in silicon have long coherence times and are a promising qubit platform.
However, electric field noise in semiconductors poses a challenge for most single- and multi-qubit operations in quantum-dot spin qubits.
We investigate the dependence of low-frequency charge noise spectra on temperature and aluminum-oxide gate dielectric thickness in Si/SiGe quantum dots with overlapping gates. 
We find that charge noise increases with aluminum oxide thickness.
We also find strong dot-to-dot variations in the temperature dependence of the noise magnitude and spectrum. 
These findings suggest that each quantum dot experiences noise caused by a distinct ensemble of two-level systems, each of which has a non-uniform distribution of thermal activation energies.
Taken together, our results suggest that charge noise in Si/SiGe quantum dots originates at least in part from a non-uniform distribution of two-level systems near the surface of the semiconductor.
\end{abstract}

\pacs{Valid PACS appear here}
\maketitle

Electron spins in silicon quantum dots are a promising platform for quantum computation~\cite{veldhorst2014addressable, tyryshkin2012electron, morello2010single, yoneda2018quantum, eng2015isotopically, maune2012coherent, watson2018programmable, veldhorst2015two}.
Long coherence times enable high fidelity qubit operations required for universal quantum computing.
Although silicon qubits largely avoid nuclear spin noise, charge noise in the semiconductor still limits both single- and multi-qubit gate fidelities. 
Moreover, charge noise levels appear to be similar in different silicon devices and materials~\cite{freeman2016comparison, petit2018spin, harvey2017coherent, shi2013coherent, wu2014two, mi2018landau, yoneda2018quantum, fogarty2018integrated}.
Because noise mitigation strategies such as device engineering, dynamical decoupling~\cite{viola1999dynamical, biercuk2009optimized}, and dynamically corrected gates~\cite{wang2012composite} rely on a detailed understanding of the noise, a thorough characterization of charge noise is essential. 

Here, we characterize the low-frequency charge noise in Si/SiGe quantum dots with overlapping gates~\cite{zajac2015reconfigurable, angus2007gate}.
We investigate the dependence of the charge noise spectrum on temperature and Al\textsubscript{2}O\textsubscript{3} gate-oxide thickness. 
We generally find that the noise increases with the aluminum oxide thickness.
Although on average the noise follows a $1/f$ power-law with a linear temperature dependence, we find strong dot-to-dot variations in the noise spectrum.
As we discuss below, we suggest that each quantum dot experiences noise caused by an ensemble of two-level systems (TLSs).
Furthermore, we suggest that separate quantum dots experience noise caused by different TLS ensembles, each of which have a different and non-uniform distribution of thermal activation energies.
In turn, variations in the TLS ensembles between dots give rise to the dot-to-dot variations in the noise.
Specifically, we analyze our measurements in the context of the Dutta-Horn (D-H) model~\cite{dutta1979energy} which considers noise generated by an non-uniform distribution of TLSs, and we find good qualitative agreement with our data.
In light of these findings, we conclude that charge noise in Si/SiGe quantum dots is caused, at least in part, by a non-uniform distribution of two-level systems near the surface of the semiconductor.

\begin{figure}[t!]
\includegraphics[width=0.48\textwidth]{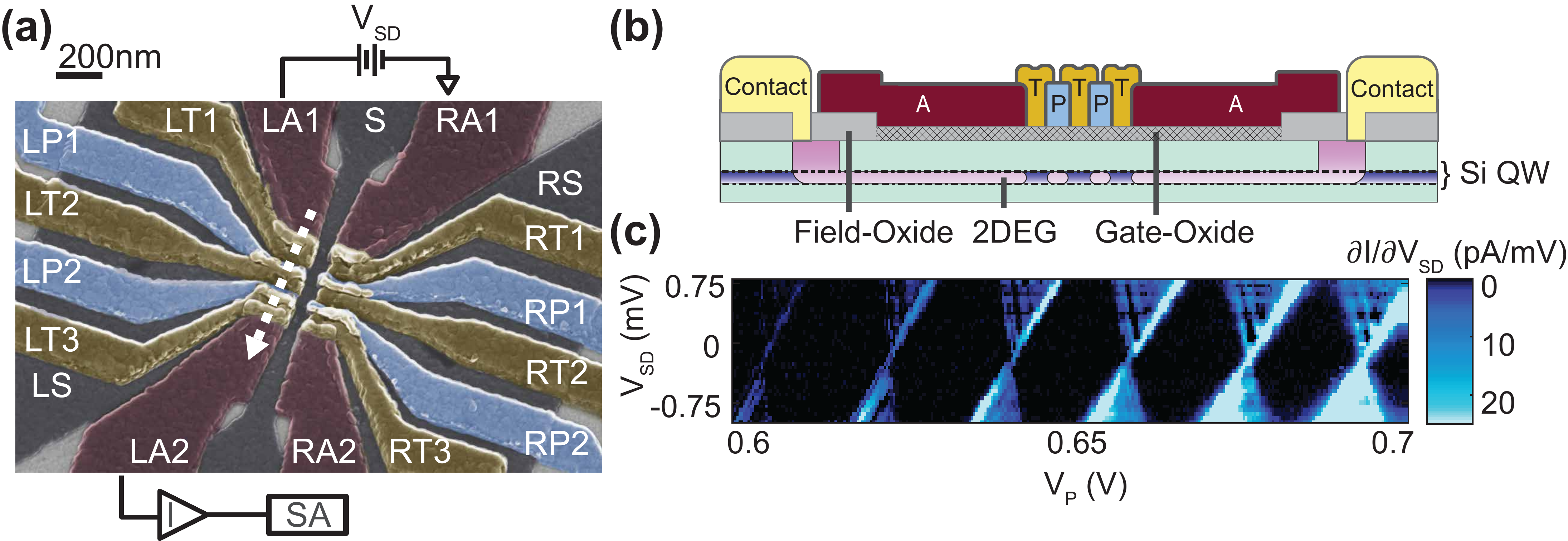}
\caption*{
\textbf{Fig. 1} Experimental setup.
\textbf{a} False color scanning electron microscope image of an overlapping-gate quantum-dot device identical to those measured. 
Dots are formed under plunger gates (blue) using screening gates (dark grey) and tunneling gates (yellow) to confine the electrons in the underlying two-dimensional electron gas (2DEG). 
A source drain bias $V_{SD}$ is applied to the 2DEG to drive a current (dashed white arrow) through the dot. 
Current is measured with a current preamplifier and spectrum analyzer (SA). 
\textbf{b} Cross section of the device along the current path.
The gate-oxide consists of Al\textsubscript{2}O\textsubscript{3} grown by atomic layer deposition.
\textbf{c} Differential conductance $\partial I/ \partial V_{SD}$ showing representative Coulomb blockade diamonds.}
\label{fig:apparatus}
\end{figure}

Devices are fabricated on an undoped Si/SiGe heterostructure with an 8~nm thick Si quantum well approximately 50~nm below the surface and a 4~nm Si cap, which forms a thin native SiO\textsubscript{2} layer on its surface. 
Voltages applied to three layers of electrostatically isolated overlapping aluminum gates defined with electron beam lithography accumulate and confine electrons in the Si quantum well forming the quantum dots [Fig. 1(a), (b)]~\cite{zajac2015reconfigurable, angus2007gate}.

Prior to quantum-dot fabrication, we deposit Al\textsubscript{2}O\textsubscript{3} on the entire wafer surface via atomic layer deposition.
On certain devices, we remove some or all of the Al\textsubscript{2}O\textsubscript{3} in the device region, allowing us to adjust the thickness of the gate dielectric.
Table I shows the parameters of the devices used here.
On Device 1, we nominally removed all of the Al\textsubscript{2}O\textsubscript{3} with H\textsubscript{3}PO\textsubscript{4} (Transene Transetch-N), which selectively etches Al\textsubscript{2}O\textsubscript{3} compared with SiO\textsubscript{2}.
We did not attempt to modify the native SiO\textsubscript{2} layer.
We also note that deposition of aluminum gates directly on an Al\textsubscript{2}O\textsubscript{3} or SiO\textsubscript{2} surface leads to interfacial layers of AlO\textsubscript{x} and modification of the underlying oxide~\cite{lim2009electrostatically,spruijtenburg2018fabrication}.
It is therefore likely that a few nm of additional AlO\textsubscript{x} exists underneath the aluminum gates on all devices, including Device 1 where we remove the deposited Al\textsubscript{2}O\textsubscript{3} by wet etch. 
In the following, we will refer to the deposited Al\textsubscript{2}O\textsubscript{3} layer that exists over the device region as the gate-oxide [Fig. 1(b)].
We measured the gate-oxide thicknesses with a combination of white-light optical reflectometry, contact profilometry, and atomic force microscopy.
See Supplemental Material~\cite{connors2019lfcnsupp} for further device fabrication details.

All devices are cooled in a dilution refrigerator with a base temperature of approximately 50~mK and then tuned to the Coulomb blockade regime [Fig 1(c)].
We apply a filtered source-drain bias of less than 1~mV across the device and measure the current $I$ with a SR570 low-noise current preamplifier.
Current noise spectra are acquired on a SR760 spectrum analyzer with the plunger gate voltage $V_P$ set on the left, right, and top of a transport peak, as well as in the Coulomb blockade regime where $I=0$ [Fig 2(a)].
We observe that the current fluctuations are most pronounced on the sides of the peak, where $\left| dI/dV_P \right|$ is largest, indicating that electrochemical potential fluctuations make the dominant contribtion to current noise~\cite{freeman2016comparison, jung2004background}.

\begin{figure}[t!]
\includegraphics[width=0.48\textwidth]{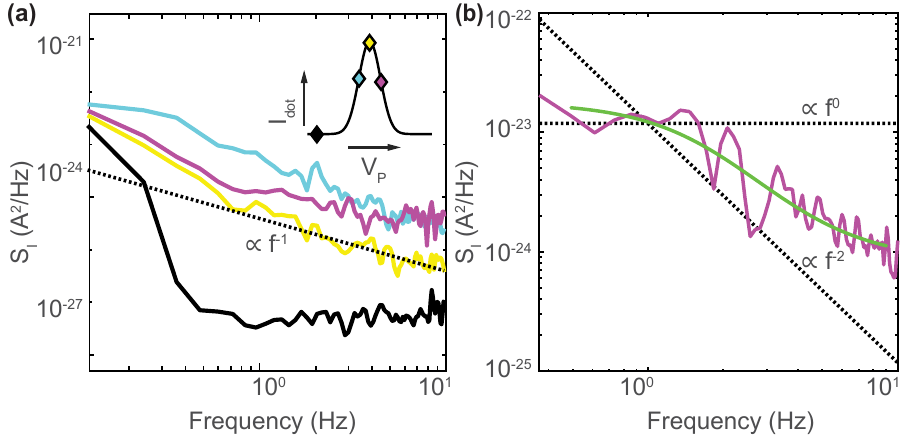}
\caption*{
\textbf{Fig. 2} Noise spectrum measurement.
\textbf{a} Current noise power spectral density measurement. 
Charge noise is measured by acquiring current noise spectra with the plunger gate $V_P$ set on both sides of a transport peak where $\left| dI/dV_P \right|$ is large as shown in the inset. 
Additional spectra are measured with $V_P$ set within the Coulomb blockade region for a baseline measurement of our experimental setup, and with $V_P$ set on top of the peak where $\left| dI/dV_P \right|$ is small as checks to ensure the measurement is sensitive to charge noise.
A dashed trendline proportional to $f^{-1}$ is shown.
\textbf{b} Measured current noise spectum showing non-power-law behavior (magenta).
The green line is a fit of the measured data to a function of the form $\frac{A}{f^{\beta}} + \frac{B}{\sfrac{f^{2}}{f_0^2}+1}$, where $A$, $B$, $\beta$, and $f_0$ are fit parameters.
Dashed lines proportional to $f^0$ and $f^{-2}$ are shown.
}\label{fig:spec+sf}
\end{figure}

In the regime where chemical potential fluctuations dominate the current noise, small current fluctuations $\delta I$ are given by

\begin{equation}\label{eq:currentnoise}
\delta I = \frac{dI}{dV_P}\frac{\delta_{\epsilon}}{\alpha},
\end{equation}

\noindent
where $\delta\epsilon$ is a small change in the electrochemical potential and $\alpha$ is the lever arm.
We extract $dI/dV_P$ from a fit of the transport peak and use Equation \ref{eq:currentnoise} to convert the acquired current noise spectrum $S_I$ to a charge noise spectrum $S_{\epsilon}$ via the relationship

\begin{equation}\label{eq:chargenoise}
S_{\epsilon} = \frac{\alpha^2S_I}{\left| \sfrac{dI}{dV_P} \right|^2}.
\end{equation}

\noindent
Note that Equation \ref{eq:currentnoise} applies only when $\delta\epsilon \ll \Delta\epsilon$, where $\Delta\epsilon$ is the width of the transport peak~\cite{beenakker1991theory}.
When $\delta\epsilon \approx \Delta\epsilon$, Equation \ref{eq:chargenoise} underestimates the charge noise.
Based on simulations, we estimate that measured values of $S_\epsilon$ differ from actual values by at most a factor of approximately 1.4 at low temperature (see Supplemental Material~\cite{connors2019lfcnsupp}).
Lever arms are extracted from Coulomb diamond measurements [Fig. 1(c)] and, when possible, confirmed from a fit of the transport peak width versus temperature.
Lever arms range from 0.036 to 0.092 eV/V, with smaller lever arms corresponding to quantum dots in devices with more gate-oxide (see Supplemental Material~\cite{connors2019lfcnsupp}).

Generally, the measured charge noise spectra have a power-law frequency dependence [Fig. 2(a)]. 
However, some spectra are better described by the sum of a power-law and a Lorentzian [Fig 2(b).]
Spectra of this type are observed on all devices. 
As we discuss below, Lorentzian noise spectra suggest the presence of individual or small numbers of TLSs.

We fit the measured $S_{\epsilon}(f)$ to a function of the form $\frac{A}{f^{\beta}} + \frac{B}{\sfrac{f^{2}}{f_0^2}+1}$, which is the sum of a power law and a Lorentzian, from 0.5-9~Hz.
Here $A$, $B$, $\beta$, and $f_0$ are fit parameters.
From this fit we directly extract the charge noise at 1~Hz, $S_{\epsilon}^{1/2}(1~\text{Hz})$, and also obtain the frequency exponent at 1~Hz, $\gamma = -\partial \ln{S_{\epsilon}}/\partial \ln{f}\left.\right|_{f=\text{1~Hz}}$, by differentiating the fit at 1~Hz. 
In total, we measured noise spectra on quantum dots on three separate devices with gate-oxide thicknesses of 0~nm, 15~nm, and 46~nm.
At the base temperature of our dilution refrigerator, we measured three quantum dots on each device to find $S_{\epsilon}^{1/2}(1~\text{Hz})$ to be $0.84 \pm 0.04$~$\mu \text{eV}/\sqrt{\text{Hz}}$ on Device 1 (0~nm gate-oxide), $0.94 \pm 0.18$~$\mu \text{eV}/\sqrt{\text{Hz}}$ on Device 2 (15~nm gate-oxide), and $1.77\pm 0.09$~$\mu \text{eV}/\sqrt{\text{Hz}}$ on Device 3 (46~nm gate-oxide).
A compilation of device parameters and charge noise values is given in Table I.
At base temperature, the charge noise generally increases with the gate-oxide thickness.
We discuss this observation further below.

\begin{table}[t!]
\begin{ruledtabular}
\begin{tabular}{*6c}
\scriptsize{Device} & \scriptsize{Gate-Oxide~(nm)} & \scriptsize{QD} & \scriptsize{$\alpha$~(eV/V)} & \multicolumn{2}{c}{\scriptsize{$S_{\epsilon}^{1/2}(1~\text{Hz})$~($\mu$eV/$\sqrt{Hz}$)}} \\
& & & & \scriptsize{QD Avg} & \scriptsize{Device Avg} \\
\hline
\hline
& & R1 & 0.088* & 0.89 &  \\
1 & 0 & L1 & 0.092* & 0.77 & 0.84 $\pm$ 0.04 \\
& & L2 & 0.070* & 0.87 & \\
\hline
& & R1 & 0.073* & 1.04 & \\
2 & 15 & L1 & 0.080* & 0.59 & 0.93 $\pm$ 0.18 \\
& & L2 & 0.073* & 1.17 & \\
\hline
& & R1 & 0.048 & 1.87 & \\
3 & 46 & R2 & 0.036 & 1.84 & 1.77 $\pm$ 0.09 \\
& & L1 & 0.038 & 1.59 & \\
\end{tabular}
\end{ruledtabular}
\caption*{
\textbf{Table I} 
Table of parameters of devices measured at base temperature of our dilution refrigerator. 
Quantum dots are specified by the plunger gate which they exist beneath as shown in Figure 1(a). 
For example, QD R1 is formed underneath plunger gate RP1.
Values of $\alpha$ labeled with an asterisk (*) have been verified via a fit of the transport peak width versus temperature.
Values for $S_{\epsilon}^{1/2}(1~\text{Hz})$ are given for both individual dots and for each device.
The reported value of the average $S_{\epsilon}^{1/2}(1~\text{Hz})$ at each dot is calculated by averaging all measurements taken on that respective dot at base temperature of the dilution refrigerator.
}
\end{table}

We investigate the temperature dependence of the charge noise at 1 Hz by sweeping the sample temperature from 50~mK to 1~K in fine-grained step sizes ranging from 2-10~mK, with larger step sizes used at higher temperatures. 
Temperature sweeps were conducted on three quantum dots on Device 1, two quantum dots on Device 2, and one quantum dot on Device 3. 
Figure 3(a) shows the average temperature dependence of $S_{\epsilon}(1~\text{Hz})$ for devices with varying gate-oxide thicknesses, averaged across all quantum dots measured on each device.
Again, we see that the noise increases with the gate-oxide thickness, especially at high temperature.
Figure 3(b) shows the spread in $\gamma(\text{1~Hz})$ as a function of temperature.
See Supplemental Material~\cite{connors2019lfcnsupp} for a detailed description of the analysis of the measured data.

Although on average, the charge noise is approximately $1/f$ and varies approximately linearly with temperature, there is significant dot-to-dot variation in the temperature dependence of $S_{\epsilon}$ and $\gamma$.
This is especially pronounced at low temperatures.
In some cases, the temperature dependence of the noise experienced by a quantum dot differs depending on which side of a transport peak $V_P$ is set [Fig. 4(b)-(c)].

It is generally thought that $1/f$ noise in semiconductors results from a distribution of bistable charge states.
Such fluctuators are regularly observed in various solid state platforms~\cite{beaudoin2015microscopic, buizert2008n, kurdak1997resistance, black1978interaction, agarwal2013polaronic, simmonds2004decoherence, lisenfeld2010measuring, grabovskij2012strain, dubois2013delocalized, hauck2014locating, pioro2005origin, jung2004background}.
A simple model for TLSs causing $1/f$ noise proposed by McWhorter~\cite{mcwhorter1957semiconductor} considers a distribution of TLSs, each of which switches between two states and contributes a Lorentzian power spectrum to the overall noise spectrum.
Under the assumption that $\tau$, the switching time of the TLSs, is thermally activated such that $\tau=\tau_0e^{E/k_BT}$, the spectrum of a single TLS is given by

\begin{equation}\label{eq:svtemp}
s_{\epsilon}(f,T) = \frac{\tau_0e^{E/k_BT}}{4\pi^2f^2\tau_0^2e^{2E/k_BT}+1}.
\end{equation}

\noindent
Here, $E$ and $\tau_0$ are the activation energy and the characteristic attempt time of the TLS, respectively.
The total power spectrum, $S_{\epsilon}(f,T)$, is simply the integral of all TLS power spectra over a distribution of activation energies $D(E)$

\begin{equation}\label{eq:Sv_int}
S_{\epsilon}(f,T) = \int \frac{\tau_0e^{E/k_BT}}{4\pi f^2\tau_0^2e^{2E/k_BT}+1} D(E)dE.
\end{equation}

\noindent
If the distribution of TLS activation energies, $D(E)$, is constant, one arrives at a total noise spectrum that is proportional to $k_BT/f$~\citep{mcwhorter1957semiconductor}.

\begin{figure}[t!]
\begin{center}
\includegraphics[width=0.48\textwidth]{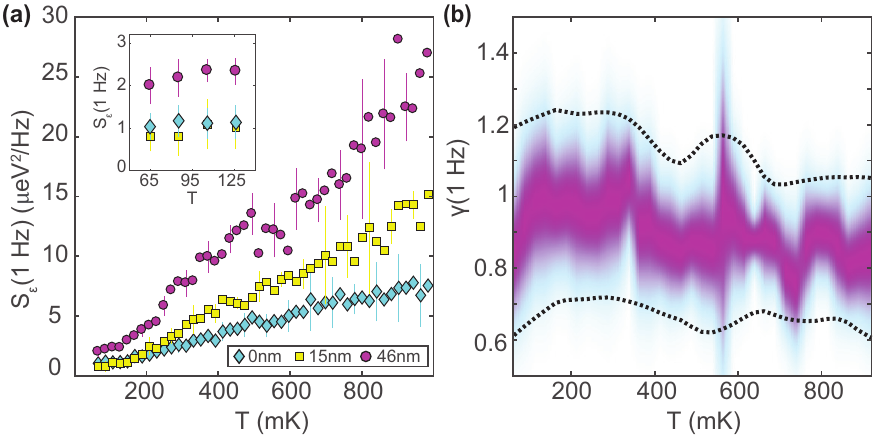}
\caption*{
\textbf{Fig. 3} 
Temperature dependence of the charge noise.
\textbf{a} Plot of the averaged $S_{\epsilon}(1~\text{Hz})$ versus temperature across three different samples with 0~nm, 15~nm, and 46~nm of gate-oxide. 
Measurements of $S_{\epsilon}(1~\text{Hz},T)$ were made on three quantum dots on the Device 1, two quantum dots on Device 2, and one quantum dot on Device 3. 
The inset shows the same data near base temperature of the dilution refrigerator. 
The data show a clear trend of increasing noise with gate-oxide thickness, especially at high temperature. 
Error bars are included on every third point and represent the standard error in the mean.
\textbf{b}
Plot of the spread of $\gamma(\text{1~Hz},T)$ of all measured samples.
The colored shadow represents the distribution of $\gamma(\text{1~Hz})$ at a given temperature, and is centered about the mean value.
The dashed lines indicate one standard deviation above and below the mean. 
}\label{fig:nvt}
\end{center}
\end{figure}

As shown in Figure 4, however, data from individual dots show strong deviations from a $1/f$ spectrum and linear temperature dependence.
Non-linear temperature dependence and anomalous frequency dependence of the charge noise have also previously been observed in semiconductor quantum dots~\cite{gungordu2019indications,dial2013charge,petit2018spin}.
In the following, we describe how a non-uniform distribution of TLSs can give rise to this behavior.
One might generally expect a non-uniform distribution of TLSs to result in anomalous temperature and spectral dependence of the noise. 
Consider, for example, a single TLS with a Lorentzian noise spectrum. 
On one hand, for frequencies $f \ll 1/2\pi\tau$ the noise is white and exponentially decreases with temperature ($\tau=\tau_0e^{E/k_BT}$ is the temperature-dependent switching time).
On the other hand, for frequencies $f \gg 1/2\pi\tau$ the noise varies as $f^{-2}$ and exponentially increases with temperature.

\begin{figure*}[t!]
\includegraphics[width=0.92\textwidth]{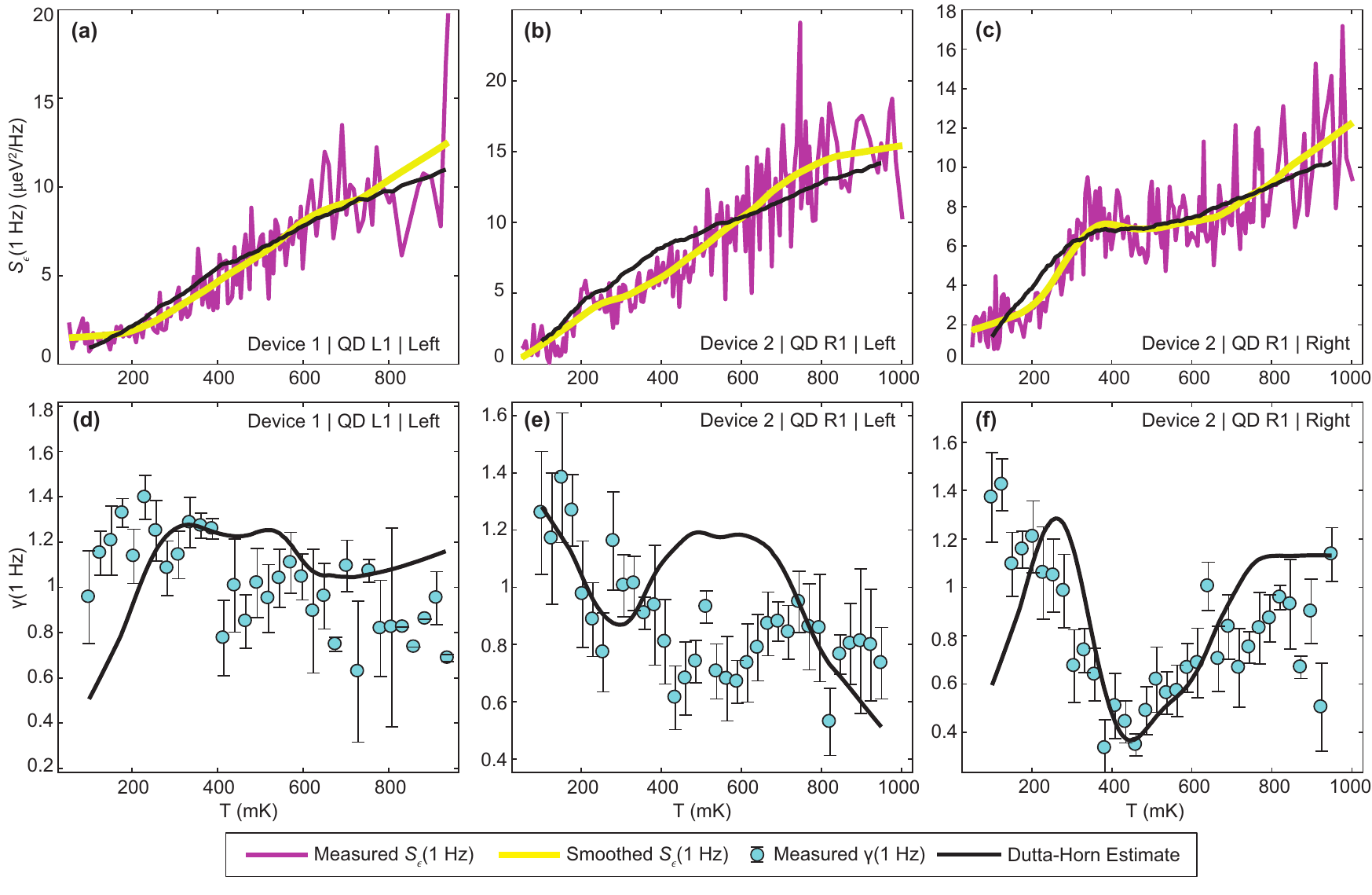}
\caption*{
\textbf{Fig. 4} Measured temperature dependence of the noise magnitude and exponent. 
\textbf{a} $S_{\epsilon}(\text{1~Hz})$ versus temperature on the left side of a transport peak for quantum dot L1 on device 1.
\textbf{b} $S_{\epsilon}(\text{1~Hz})$ versus temperature on the left side of a transport peak for dot R1 on device 2.
\textbf{c} $S_{\epsilon}(\text{1~Hz})$ versus temperature on the right side of a transport peak for dot R1 on device 2.
\textbf{d} Averaged measurements of $\gamma(\text{1~Hz})$ versus temperature on the left side of the transport peak for dot L1 on device 1.
\textbf{e} Averaged measurements of $\gamma(\text{1~Hz})$ versus temperature on the left side of the transport peak for dot R1 on device 2.
\textbf{f} Averaged measurements of $\gamma(\text{1~Hz})$ versus temperature on the right side of the transport peak for dot R1 on device 2.
The smooth yellow lines are generated by taking a moving average of the data.
The black lines in \textbf{a}, \textbf{b}, and \textbf{c} are the estimates for $S_{\epsilon}(\text{1~Hz},T)$ using Equation \ref{eq:gamma} of the Dutta-Horn model and the data in \textbf{d}, \textbf{e}, and \textbf{f}, respectively.
Error bars in \textbf{d}, \textbf{e}, and \textbf{f} represent the standard error associated with averaging the data over temperature ranges of 25~mK. 
The solid lines in \textbf{d}, \textbf{e}, and \textbf{f} are estimates for $\gamma(\text{1~Hz},T)$ using Equation \ref{eq:gamma} of the Dutta-Horn model and the smoothed yellow lines in \textbf{a}, \textbf{b}, and \textbf{c}, respectively.
} \label{fig:DH}
\end{figure*}

The model of Dutta and Horn~\cite{dutta1979energy} extends the McWhorter model to account for a non-uniform distribution of TLSs. 
The D-H model has successfully described $1/f$ noise in a large variety of solid state systems~\cite{dutta1979energy, fleetwood20151, liu2016low, fleetwood1984temperature}. 
Under the assumption that the width of the distribution of activation energies is larger than $k_BT$, one can expand the result of Equation~\ref{eq:Sv_int} in powers of $T$ to obtain

\begin{equation}\label{eq:Sv_linear}
S_{\epsilon}(f,T) =  \frac{k_BT}{2\pi f}D(\tilde{E}),
\end{equation}

\noindent 
where $\tilde{E} = -k_BT\ln{(2\pi f\tau_0)}$.
Equation \ref{eq:Sv_linear} shows that if $D(E)$ is not constant, then $S_{\epsilon}(f,T)$ will not vary linearly with temperature.
Additionally, if $\gamma \neq 1$, then Equation ~\ref{eq:Sv_linear} suggests $D(\tilde{E})$ must not be constant.
Moreover, by defining $\gamma \equiv -\partial\ln{S_{\epsilon}}/\partial\ln{f}$, one can use Equation \ref{eq:Sv_linear} to obtain the following relation between the noise power $S_{\epsilon}(f,T)$ and the frequency exponent $\gamma(f,T)$

\begin{equation}\label{eq:gamma}
\gamma(f,T) = 1 - \frac{1}{\ln{\left(2\pi f \tau_0\right)}}\left[ \frac{\partial \ln{S_{\epsilon}(f,T)}}{\partial \ln{T}} -1 \right].
\end{equation}

\noindent
Equations~\ref{eq:Sv_linear} and \ref{eq:gamma} are the basis of the D-H model.
Equation~\ref{eq:Sv_linear} relates the temperature dependence of the noise to the density of the TLSs. 
Equation~\ref{eq:gamma} relates the temperature dependence of the frequency exponent to that of the noise magnitude.
Note that Equation~\ref{eq:gamma} implies that deviations from $\gamma=1$ imply a non-uniform distribution of TLSs and a non-linear temperature dependence of $S_{\epsilon}(f,T)$ as discussed above.
Figure 4 shows representative plots of our measurements of $S_{\epsilon}(\text{1~Hz},T)$ and $\gamma(\text{1~Hz},T)$.
All data sets show deviations from both $\gamma=1$ and linear temperature dependence of $S_{\epsilon}(\text{1~Hz},T)$.

According to the D-H model, these data suggest a non-uniform distribution of activation energies $D(E)$.
We show that our data is in qualitative agreement with the D-H model in several ways. 
First, using the measurements of $\gamma(\text{1~Hz},T)$, we integrate Equation~\ref{eq:gamma} to generate a prediction for $S_{\epsilon}(\text{1~Hz},T)$.
We generally observe good qualitative agreement with our measurements of $S_{\epsilon}(\text{1~Hz},T)$ using this approach, although some of the sharp features are not perfectly captured [Fig. 4(a)-(c)].
Second, to generate a predicted form of $\gamma(\text{1~Hz},T)$ from our measured noise power spectral density, we smooth the data using a moving 50-point average.
We then take the logarithmic derivative of the smoothed line to extract a prediction for $\gamma(\text{1~Hz},T)$ based on Equation~\ref{eq:gamma} (see Supplemental Material~\cite{connors2019lfcnsupp} for details regarding generating predictions for $S_{\epsilon}(\text{1~Hz},T)$ and $\gamma(\text{1~Hz},T)$ via the D-H model).
Again, our predictions based on the D-H model show reasonable qualitative agreement with the data [Fig. 4(d)-(f)].
In all cases, we fixed the maximum attempt frequency $\omega_0 = 1/\tau_0$ at 5~$\text{s}^{-1}$ to maximize the fit quality across all data sets. 
The required value of $\omega_0$, which controls the size of deviations from $\gamma=1$ and linear temperature dependence in the D-H model, is puzzling because $1/f$ noise has been observed at higher frequencies in Si/SiGe quantum dots~\cite{mi2018landau, yoneda2018quantum}.
One possible explanation is that the assumptions of the D-H model are not entirely satisfied in our experiment. 
For example, the presence of sharp features in the activation energy distribution, as suggested by individual Lorentzian features in the measured spectra, may cause strong deviations from $\gamma=1$.
However, we note that predictions of the D-H model depend logarithmically on $\omega_0$, so our results only weakly depend on its precise value.

Figure 4 shows the predictions for $S_{\epsilon}(\text{1~Hz},T)$ and $\gamma(\text{1~Hz},T)$ made by the D-H model for three representative cases with varying quality of agreement between measurements and predictions.
Given the generally good qualitative agreement between our data and the D-H predictions, we suggest that the charge noise results from a non-uniform distribution of TLSs.
We note that the observation of Lorentzian features in the noise spectra corroborate this view [Fig 2(b)].
See Supplemental Material~\cite{connors2019lfcnsupp} for comparisons between our data and the D-H model for all devices measured.

We obtain further insight into the nature of the noise source by measuring the temporal correlation of the charge noise on two neighboring quantum dots. 
First, we tune dots L1 and R1 on Device 2 to the Coulomb blockade regime.
We set both plunger gates to the sides of their respective transport peaks, and we acquire a time series of current fluctuations on each dot simultaneously for 3200~seconds and repeat this procedure 20 times.
We calculate correlation coefficients of the current fluctuations between dots for each 3200-second time series and average the result across the 20 repetitions and find a correlation coeffecient $\rho(\delta I_{L1},\delta I_{R1}) = -0.006 \pm 0.032$, which indicates that the noise at each dot is independent and local.
See Supplemental Material~\cite{connors2019lfcnsupp} for details regarding the calculation of the correlation coefficient.
Together with our earlier results, it seems plausible that charge noise is caused by a small number of TLSs in close proximity to each quantum dot.

Our data suggest several possible explanations for the charge noise.
One explanation is that the aluminum oxide itself contains the TLSs.
In this case we would expect that reducing the oxide thickness would reduce the overall noise.
It is also possible that the TLSs exist in the semiconductor near its surface or at the SiO\textsubscript{2}/AlO\textsubscript{x} interface, and decreasing the aluminum oxide thickness improves screening effects from the metallic gates.
If the individual TLSs consist of dipole charge traps~\cite{burnett2016analysis,burin2015low,martinis2005decoherence,sarabi2016projected}, however, the metal gates will only screen dipoles oriented parallel to the surface. 
Image charges associated with dipoles oriented perpendicular to the surface would increase their contribution to the noise.
For randomly oriented dipoles, one would not expect a significant change in the noise as the distance to the metal gates decreases.
Thus, we suggest that the charge noise is caused at least in part by TLSs in the aluminum oxide, or dipole TLSs oriented parallel to the wafer surface and located either in the semiconductor near its surface or at the SiO\textsubscript{2}/AlO\textsubscript{x} interface.
However, it seems more likely that interface TLSs would be oriented parallel to the wafer surface than TLSs in the bulk of the semiconductor.
In all of these cases, we emphasize that reducing the AlO\textsubscript{x} thickness is expected to reduce the noise, as suggested by this and previous work~\cite{zimmerman2014charge}.

In summary, we find that the presence of an aluminum-oxide gate dielectric layer tends to increase charge noise in Si/SiGe quantum dots.
We observe that most quantum dots on a given device suffer from similar levels of noise, though there often exist significant dot-to-dot variations in the temperature dependence of the noise across dots in the same device.
In the context of the Dutta-Horn model, our findings suggest that a non-uniform distribution of TLSs is responsible for the charge noise.
Based on our results, it seems plausible that a small number of TLSs near the surface of the semiconductor or in the gate-oxide cause the charge noise.
Our data underscore the importance of controlling defect densities in the gate-stack on top of silicon quantum dots. 
Our results also emphasize the importance of fully characterizing the charge noise of individual quantum dots to determine optimal spin qubit dynamical decoupling protocols. 
Furthermore, we suggest the use of as little aluminum oxide as possible in the active region of Si/SiGe spin qubits as an effective means to reduce charge noise.

Research was sponsored by the Army Research Office and was accomplished under Grant Numbers W911NF-16-1-0260 and W911NF-19-1-0167.  The views and conclusions contained in this document are those of the authors and should not be interpreted as representing the official policies, either expressed or implied, of the Army Research Office or the U.S. Government.  The U.S. Government is authorized to reproduce and distribute reprints for Government purposes notwithstanding any copyright notation herein.
E.J.C. was supported by ARO and LPS through the QuaCGR Fellowship Program.

%

\newpage
\clearpage
\newpage
 
\section*{Supplemental Material}

\subsection{Device}
All devices were fabricated on an undoped Si/SiGe heterostructure, and Al\textsubscript{2}O\textsubscript{3} was deposited on the entire surface via atomic layer deposition.
The Al\textsubscript{2}O\textsubscript{3} in the active region of Device 1 was etched completely via Transene Transetch-N.
The Al\textsubscript{2}O\textsubscript{3} in the active region of Device 3 was partially etched using buffered oxide etch such that 46~nm of gate-oxide remained.
A forming gas anneal step was performed on Device 2 immediately following aluminum oxide deposition.
We measured Al\textsubscript{2}O\textsubscript{3} thicknesses in both the unetched and etched (if the device was etched) regions of each device using white-light optical reflectometry.
On Devices 1 and 3, we cross-checked these measurements by measuring the height of the etch boundary via contact profilometry and atomic force microscopy, respectively. 
Once the Al\textsubscript{2}O\textsubscript{3} gate-oxide was deposited or removed, three layers of overlapping aluminum gates were defined on all devices via a combination of photolithography and electron beam lithography steps.
Aluminum was deposited via thermal evaporation on Devices 1 and 2, and electron beam evaporation on Device 3. 
For thermally deposited aluminum layers, we created a buffer layer by depositing 3~nm of Al at a rate less than 0.1~$\text{\AA/s}$ before depositing the rest of the Al at rate of approximately 2.0~$\text{\AA/s}$.
Empirically, the buffer layer minimizes gate-to-gate leakage.
\newline

\subsection{Extraction of $\mathbf{\alpha}$}
The lever arm $\alpha$ is extracted on each quantum dot from Coulomb diamond measurements as seen in Figure 1(c) in the main text. 
Slopes of the diamonds $m_S$ and $m_D$, corresponding to the electron source and drain, respectively, are measured.
The lever arm is determined via the equation

\begin{equation}
\alpha = \left| \frac{m_Sm_D}{m_S-m_D} \right|.
\end{equation} 

\noindent
On some quantum dots, we verify this measurement of $\alpha$ using the traces of the transport peaks acquired during the temperature sweep.
We can then extract $\alpha$ by fitting the transport peaks, $I(V)$, to a function of the form

\begin{equation}
I(V) = A + B\cosh^{-2}\left(\frac{\alpha(V_0-V)}{2k_BT_e}\right),
\end{equation}

\noindent
where $A$, $B$, $V_0$, and $T_e$ are fit parameters~\cite{beenakker1991theory}. 
We then extract $\alpha$ from fitting the measured electron temperature, $T_e$, versus the mixing chamber temperature, $T$, to a straight line with unit slope and zero offset for $T > \text{500~mK}$.
In all cases, we get reasonable agreement in our determination of $\alpha$ across these two methods.
\newline

\subsection{$\mathbf{S_{\epsilon}}$ Measurement Simulation}
We measure charge noise on a quantum dot by setting the voltage of its corresponding plunger gate $V_P$ on the left and right sides of a transport peak, and measuring current noise spectra.
Since $\left| dI/dV_P \right| $ is large, fluctuations in the electrochemical potential of the dot are the dominant contribution to the current noise.
We then use Equations 1 and 2 from the main text to relate the measured current noise to electrochemical potential noise.
However, these equations are valid when $\delta\epsilon \ll \Delta\epsilon$, where $\delta\epsilon$ is the elctrochemical potential fluctuation and $\Delta\epsilon$ is the width of the transport peak. 
If $\delta\epsilon \approx \Delta\epsilon$, large fluctuations in the electrochemical potential reduce the time-averaged value of $\left| dI/dV_P \right|$.
As a result, the measured noise, $S_{\epsilon}^m$, will underestimate the real noise, $S_{\epsilon}^r$.
At high temperature the transport peak is thermally broadened such that the width is much larger than fluctuations in the electrochemical potential, but at low temperature this source of error can can become relevant.

To estimate the impact of this effect on our measurements, we simulate $1/f$ charge noise, $S_{\epsilon}^r$, for each separate spectrum acquired on both the left and right sides of the transport peak during all temperature sweeps.
We fix the magnitude of $S_{\epsilon}^r(\text{1~Hz})$ at the measured value for the corresponding acquired spectrum.
The noise is simulated in a bandwidth of $f_1=0.12$~Hz to $f_2=60$~Hz, where $f_1$ corresponds to the total acquisition time per spectrum and $f_2$ corresponds to the sampling rate in our measurements of the transport peak shape. 
We simulate the resulting current noise using fits of the measured transport peaks. 
We convert this simulated current noise back to charge noise, $S_{\epsilon}^m$, using Equations 1 and 2 from the main text.
We plot the ratio $S_{\epsilon}^r(1~\text{Hz})/S_{\epsilon}^m(1~\text{Hz})$ in Supplementary Figure 1. 
Based on our simulations, we expect that the measured noise power differs from the real value by at most a factor of 1.4 at low temperature.

\begin{figure*}[b!]
\includegraphics[width=0.8\textwidth]{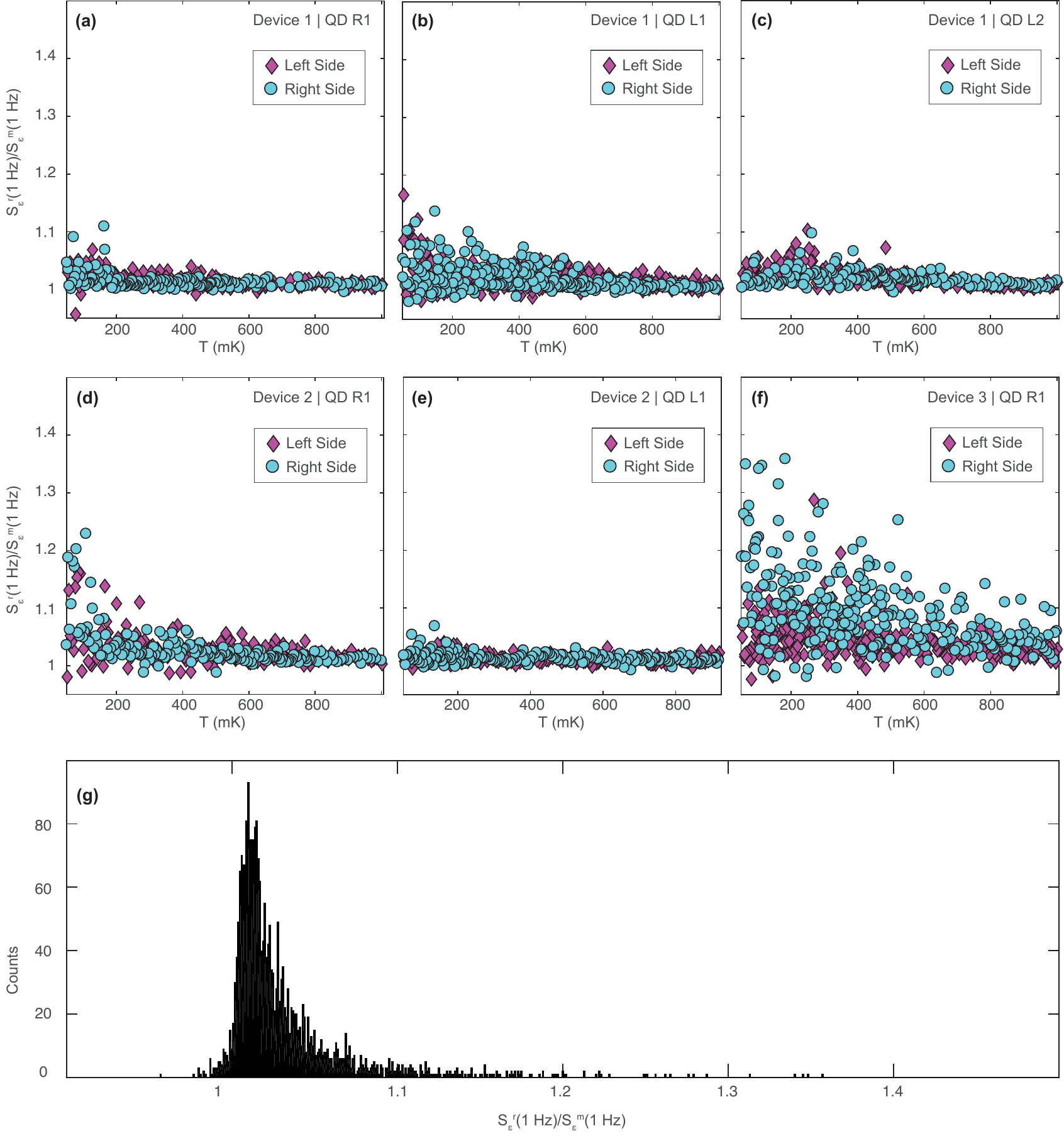}\label{fig:ext1}
\caption*{
\textbf{Supplementary Fig. 1}
Simulation of systematic underestimation of charge noise measurements. 
The ratio $S_{\epsilon}^{r}(\text{1~Hz})/S_{\epsilon}^{m}(\text{1~Hz})$, as a function of temperature, is shown for all measurements made on all quantum dots on which charge noise as a function of temperature was measured.
\textbf{a} Device 1 quantum dot R1.
\textbf{b} Device 1 quantum dot L1.
\textbf{c} Device 1 quantum dot L2.
\textbf{d} Device 2 quantum dot R1.
\textbf{e} Device 2 quantum dot L1.
\textbf{f} Device 3 quantum dot R1.
\textbf{g} Histogram of all values of $S_{\epsilon}^{r}(\text{1~Hz})/S_{\epsilon}^{m}(\text{1~Hz})$ shown in \textbf{a}-\textbf{f}.
}
\end{figure*}


\subsection{Analysis of $\mathbf{S_{\epsilon}^{1/2}(1~\text{Hz})}$ and $\gamma(\text{1~Hz})$}
For each device measured, current noise spectra are acquired on a SR760 spectrum analyzer with the plunger gate voltage $V_P$ set on the left, right, and top of a transport peak, as well as in the Coulomb blockade regime where $I=0$.
A minimum of 10 spectra are acquired at each position.
Each spectrum contains 400 points from 0.1224 to 48.96~Hz.
Spectra acquired on the top of the transport peak and in the Coulomb blockade regime are used only to verify that electrochemical potential fluctuations make the dominant contribtion to current noise for each measurement, and are not used in the analysis of the charge noise.
We extract $dI/dV_P$ from a fit of the transport peak and use Equation 2 of the main text to convert the current noise spectrum acquired on the left and right sides of the transport peak to charge noise spectrum.
We fit the measured charge noise spectrum, $S_{\epsilon}(f)$, to a function of the form $\frac{A}{f^{\beta}} + \frac{B}{\sfrac{f^{2}}{f_0^2}+1}$ from 0.5-9~Hz, where $A$, $B$, $\beta$, and $f_0$ are fit parameters.
From this fit we directly extract the charge noise at 1~Hz, $S_{\epsilon}^{1/2}(1~\text{Hz})$, and also obtain the frequency exponent at 1~Hz, $\gamma = -\partial \ln{S_{\epsilon}}/\partial \ln{f}\left.\right|_{f=\text{1~Hz}}$, by differentiating the fit at 1~Hz. 
All data reported in this work pertaining to a particular quantum dot are acquired from the same transport peak on that dot. 

The average value of $S_{\epsilon}^{1/2}(1~\text{Hz})$ at base temperature for each quantum dot is reported in Table I of the main text, and is calculated by averaging all values of $S_{\epsilon}^{1/2}(1~\text{Hz})$ extracted from spectra acquired at base temperature of the dilution refrigerator on a particular quantum dot.
In all but three cases, the reported value is determined by averaging $S_{\epsilon}^{1/2}(1~\text{Hz})$ extracted from 20 measured spectra (10 spectra on each side of the transport peak). 
On three separate dots we made a large number of measurements of the charge noise spectra at base temperature, and the reported value of $S_{\epsilon}^{1/2}(1~\text{Hz})$ in Table I is the average value of $S_{\epsilon}^{1/2}(1~\text{Hz})$ extracted from all measured spectra.
On dot R1 on Device 2, a total of 7780 spectra were acquired.
On dot L1 on Device 2, a total of 6280 spectra were acquired.
On dot L1 on Device 3, a total of 920 spectra were acquired. 
The standard deviations of the acquired values of $S_{\epsilon}^{1/2}(1~\text{Hz})$ across these measurements were 0.026, 0.075, and 0.138~$\mu \text{eV}/\sqrt{\text{Hz}}$, respectively.
As the standard deviation of these measurements is much less than the dot to dot variation of $S_{\epsilon}^{1/2}(1~\text{Hz})$, we report the average $S_{\epsilon}^{1/2}(1~\text{Hz})$ of the device as the mean of the average value of each dot. 

We investigate the temperature dependence of the charge noise at 1 Hz by sweeping the sample temperature from 50~mK to 1~K in fine-grained step sizes ranging from 2-10~mK, with larger step sizes used at higher temperatures. 
At each temperature, 10 spectra are aquired on each side of a transport peak.
Temperature sweeps were conducted on three quantum dots on Device 1, two quantum dots on Device 2, and one quantum dot on Device 3. 
Data shown in Figure 3(a) of the main text are downsampled by averaging together all measured values of $S_{\epsilon}^{1/2}(1~\text{Hz})$ acquired at temperatures within 20~mK ranges, including measurements on both sides of a transport peak, as well as measurements made on different quantum dots on the same device. 
Error bars represent the standard error of the mean in the downsampled data. 
To analyze our measurements of $\gamma(\text{1~Hz})$, we determine the mean value of $\gamma(\text{1~Hz})$ as a function of temperature across all quantum dots on all devices measured by averaging together all measured values of $\gamma(\text{1~Hz})$ acquired at temperatures within 25~mK ranges.
The spread of $\gamma(\text{1~Hz})$ is simply the standard deviation of these data.
These data are shown in Figure 3(b) of the main text. 
\newline

\subsection{Dutta-Horn Model}
Extended Data Figure 2 shows measurements of $S_{\epsilon}(\text{1~Hz},T)$ and $\gamma (\text{1~Hz},T)$ on all quantum dots, as well as the predictions made using Equation 6 in the main text.
Measured values of $S_{\epsilon}(\text{1~Hz})$ and $\gamma (\text{1~Hz})$ are obtained from a fit of the measured spectra to a function of the form $\frac{A}{f^{\beta}}+\frac{B}{\sfrac{f^2}{f_0^2}+1}$ from 0.5-9~Hz.

Dutta-Horn predictions for $S_{\epsilon}(\text{1~Hz},T)$ are made from the measured values of $\gamma (\text{1~Hz},T)$ via integrating Equation 6.
In most cases, D-H predictions of $S_{\epsilon}(\text{1~Hz},T)$ capture the broad features in the data.

We make predictions for $\gamma (\text{1~Hz},T)$ using the measured values of $S_{\epsilon}(\text{1~Hz},T)$ with Equation 6.
We smooth $S_{\epsilon}(\text{1~Hz},T)$ by taking a moving 50-point average, and then we take the logarithmic derivative of the smoothed line. 
Most temperature sweeps are done in 2-5~mK steps, so a 50-point average results in a smoothing of points across a range of approximately 250~mK, possibly explaining why the predictions made by the D-H model often miss narrow features in the data.

\begin{figure*}[htb!]
\includegraphics[width=0.87\textwidth]{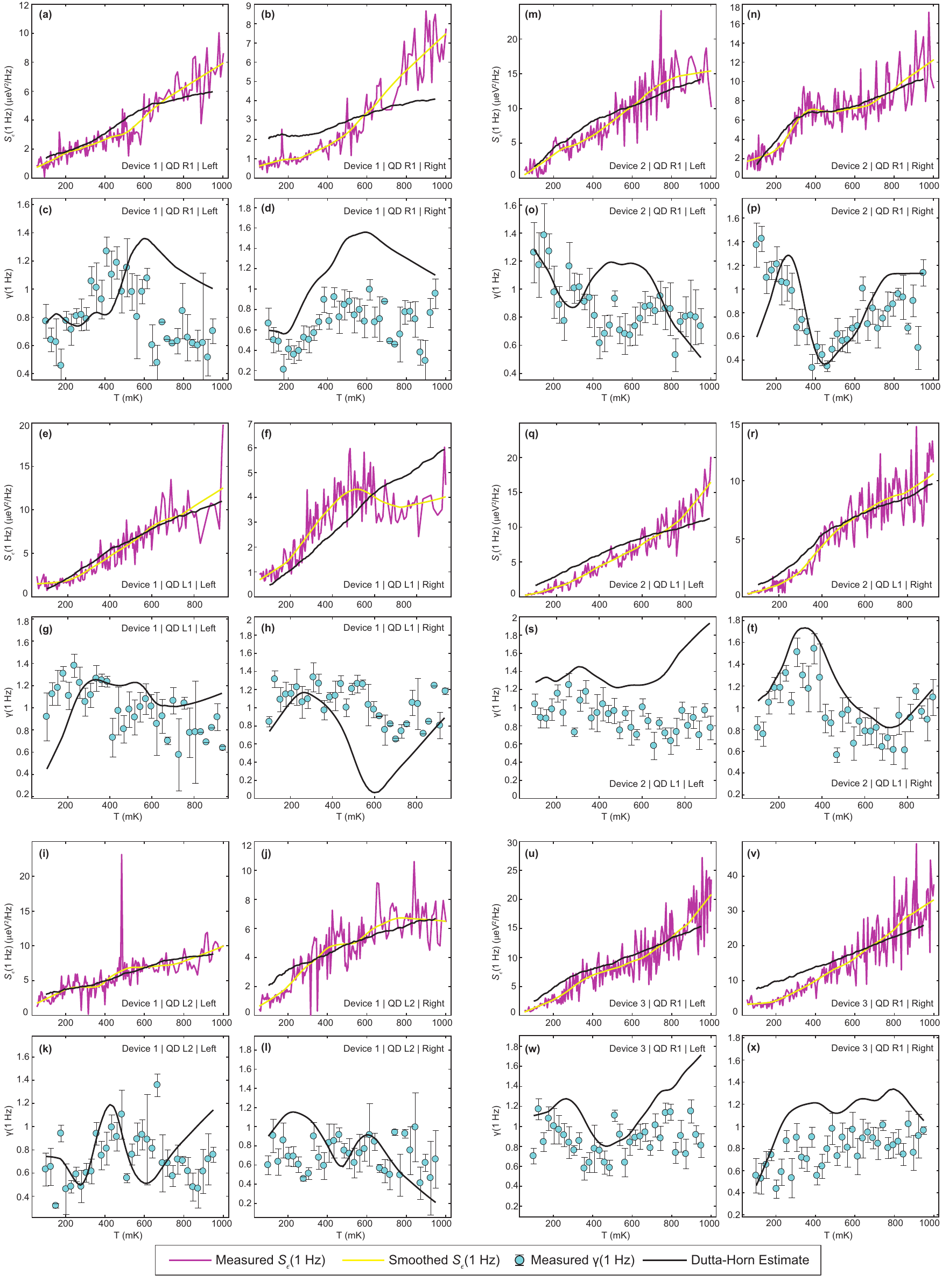}\label{fig:ext2}
\caption*{
\textbf{Supplementary Fig. 2}
Measured $S_{\epsilon}(\text{1~Hz},T)$ and $\gamma(\text{1~Hz},T)$ of all quantum dots measured, and their corresponding Dutta-Horn predictions.
\textbf{a-d} Device 1 quantum dot R1.
\textbf{e-h} Device 1 quantum dot L1.
\textbf{i-l} Device 1 quantum dot L2.
\textbf{m-p} Device 2 quantum dot R1.
\textbf{q-t} Device 2 quantum dot L1.
\textbf{u-x} Device 3 quantum dot R1.
}
\end{figure*}

\subsection{Temporal Correlation Measurement}
In order to measure the temporal correlation of the charge noise, we tune gate voltages on Device 2 such that quantum dots L1 and R1 are formed approimaxtely 100~nm apart underneath plunger gates LP1 and RP1, respectively.
Because the dots are separated by a screening gate there is negligible interdot tunneling.
We separately bias each dot and adjust their respective tunnel barriers such that the peak current is a few hundred pA.
We set each plunger gate voltage to the side of a transport peak.
We then simultaneously acquire 3200-second time series of the current fluctuations through the each device at a sampling rate of 1~kHz, and we repeat this until a total of 20 time series have been acquired from each dot.
Our data acquisition card has spurious peaks at multiples of 0.33~Hz and multiples 0.5~Hz.
Thus, we digitally apply a low-pass filter by first calculating the FFT of each time series, multiplying each of the resulting spectra by a transfer function of the form $f(\omega) = 1/(1+i\omega/\omega_C)$ and then taking the inverse Fourier transform of the filtered spectra. Here, $\omega_C/2\pi=0.3~\text{Hz}$ is the cutoff frequency of the low-pass filter.
We determine the correlation coefficient 

\begin{equation}
\rho(A,B) = \frac{\text{cov}(A,B)}{\sigma_{A}\sigma_{B}}
\end{equation}

\noindent
for each of the 20 pairs of simultaneously acquired time series and average correlation coefficients across all pairs to get an average correlation coefficient of $\rho(\delta I_{L1},\delta I_{R1}) = -0.006 \pm 0.032$.

\end{document}